# actifpTM: a refined confidence metric of AlphaFold2 predictions involving flexible regions

Julia K. Varga[1], Sergey Ovchinnikov[2] and Ora Schueler-Furman[1]*

[1]Department of Microbiology and Molecular Genetics, Institute for Biomedical Research Israel-Canada, Faculty of Medicine, The Hebrew University of Jerusalem, Jerusalem 9112001, Israel
[2]Department of Biology, Massachusetts Institute of Technology, Cambridge, MA 02139

* Correspondence: ora.furman-schueler@mail.huji.ac.il

ORCID-s:
Julia K. Varga: 0000-0002-5318-2798
Sergey Ovchinnikov:  0000-0003-2774-2744
Ora Schueler-Furman: 0000-0002-1624-0362

## Abstract

One of the main advantages of deep learning models of protein structure, such as Alphafold2, is their ability to accurately estimate the confidence of a generated structural model, which allows us to focus on highly confident predictions.The ipTM score provides a confidence estimate of interchain contacts in protein-protein interactions. However, interactions, in particular motif-mediated interactions, often also contain regions that remain flexible upon binding. These non-interacting flanking regions are assigned low confidence values and will affect iPTM, as it considers all interchain residue pairs, and two models of the same motif-domain interaction, but differing in the length of their flanking regions, would be assigned very different values. Here we propose actifpTM (actual interface pTM), a modified ipTM measure, that focuses on the confident region of an interaction, resulting in a more robust measure of interaction confidence, even when not the full interaction is structured. actifpTM has been incorporated into ColabFold.

# Introduction

Since 2020, AlphaFold2 (AF2, (Evans *et al.*, 2021; Jumper *et al.*, 2021) and other similar deep learning algorithms (Ahdritz *et al.*, 2024; Baek *et al.*, 2023) have revolutionized structural biology, by making swift and confident predictions of millions of protein structures and complexes available. Their appearance was also groundbreaking in terms of scoring of generated models, as they were trained to be able to assess the confidence of the predictions, along with the output. One of the scores introduced in AF2-Multimer that is used for assessing complex confidence is ipTM (interface predicted template modeling score), which evaluates the predicted relative confidence of binding partners (Evans *et al.*, 2021; Jumper *et al.*, 2021).

ipTM (and a confidence metric, defined as a weighted sum of ipTM and pTM) is a frequently used score for assessing docked peptide-protein models (Teufel *et al.*, 2023; Lee *et al.*, 2024; Bret *et al.*, 2024), but it was trained on PDB structures without long intrinsically disordered regions or flexible flanking regions of peptides, since these are usually not resolved in crystal structures. Although it is called an interface pTM, it is actually an inter*chain* pTM score, as it takes into account the whole length of all binding partners with equal weight. Therefore, the score changes significantly if a sequence that includes also non-structured flanking regions is provided (e.g., by including disordered regions to extend a defined domain or peptide motif, see **Figure 1A**), making comparison between otherwise similarly well modeled complexes challenging. This phenomenon particularly affects peptide-protein-like interactions as their binding interface regions are very short.

A seemingly obvious solution to the problem would be to run predictions on minimal binding regions without the flanking sequences. This has however been shown to be a less effective method for modeling (Bret *et al.*, 2024; Lee *et al.*, 2024). Even though these regions are usually flexible and not necessarily confidently modeled, AF2 seems to utilize the context for making accurate predictions. Alternatively, one can drop regions based on other confidence metrics and calculate the interface predicted alignment error score (iPAE) (Kruse *et al.*, 2024; Kim *et al.*, 2024) for the confident interface residues. However, since ipTM is more sensitive than iPAE (Teufel *et al.*, 2023; Lee *et al.*, 2024; Bret *et al.*, 2024), it would be advantageous to modify ipTM directly (and consequently, the confidence score). In this study, we introduce actifpTM (<u>act</u>ual <u>i</u>nter<u>f</u>ace pTM), a modified ipTM score that focuses on confident interface residues. To do so, we first detail how ipTM is calculated, and then describe how this calculation needs to be modified to account for the relevant residues only.

ipTM is calculated based on the equation of the original TM-score (Zhang and Skolnick, 2004):

$$TM\text{-}score = max\left[\frac{1}{L_{target}} \sum_{i}^{L_{common}} \frac{1}{1+\left(\frac{d_i}{d_0(L_{target})}\right)^2}\right]$$

where $L_{common}$ is the length of the alignment, $d_0$ is a normalization factor between alignments of different length and $d_i$ is the distance between residue *i* in the target and the query. TM-align searches for the maximum of the scores of the different possible alignments. When calculating the pTM-score in AF2, the length of the alignment is taken as the length of the concatenated input sequences, which is also used to calculate $d_0$. During the calculation, the predicted aligned error head of AF2 outputs a 3D matrix with probabilities for each of the 64 error bins (*i.e.*, the probability of the error for every residue pair being 0-0.5 Å, 0.5-1 Å, … 30.5-31 Å and beyond, **Figure 1B**). For each of these bins, a maximum theoretical TM-score can be calculated by taking the center of the bin as $d_i$ and using the TM-score formula. The pairwise probability matrices of the 64 error bins are multiplied by their respective maximum theoretical TM-scores, and summed for each residue pair to obtain a 2D pairwise pTM-score matrix (**Supplementary Figure 1A**). Then a residue mask is calculated: for pTM, the whole matrix is taken into account, while for ipTM the areas that are marking intrachain contacts are masked (**Supplementary Figure 1B**). Either way, the residue mask is then row-wise normalized, and then the pairwise pTM matrix is multiplied by it. The resulting matrix is row-wise summed and then the maximum of the per-residue values are taken as the final ipTM value. This is theoretically equivalent to the model and the native structure being superimposed onto each other on every possible residue, and reporting the TM-score of the best alignment.

The above described calculations assign equal weights to each residue pair, $d_i$, regardless of them being in contact with each other or not. To overcome this issue, for actifpTM we modify the masking of the original ipTM calculations to take into account the predicted distance probabilities as residue-pair weights (de facto, shortening $L_{common}$ by skipping $d_i$ values of non-interacting residues, **Figure 1C,D)**. Consequently, the score does not depend anymore on the length of the additional regions and thus provides similar values for structural models with a bound part of similar confidence. The calculation of actifpTM can be easily integrated into the AF2 pipeline.

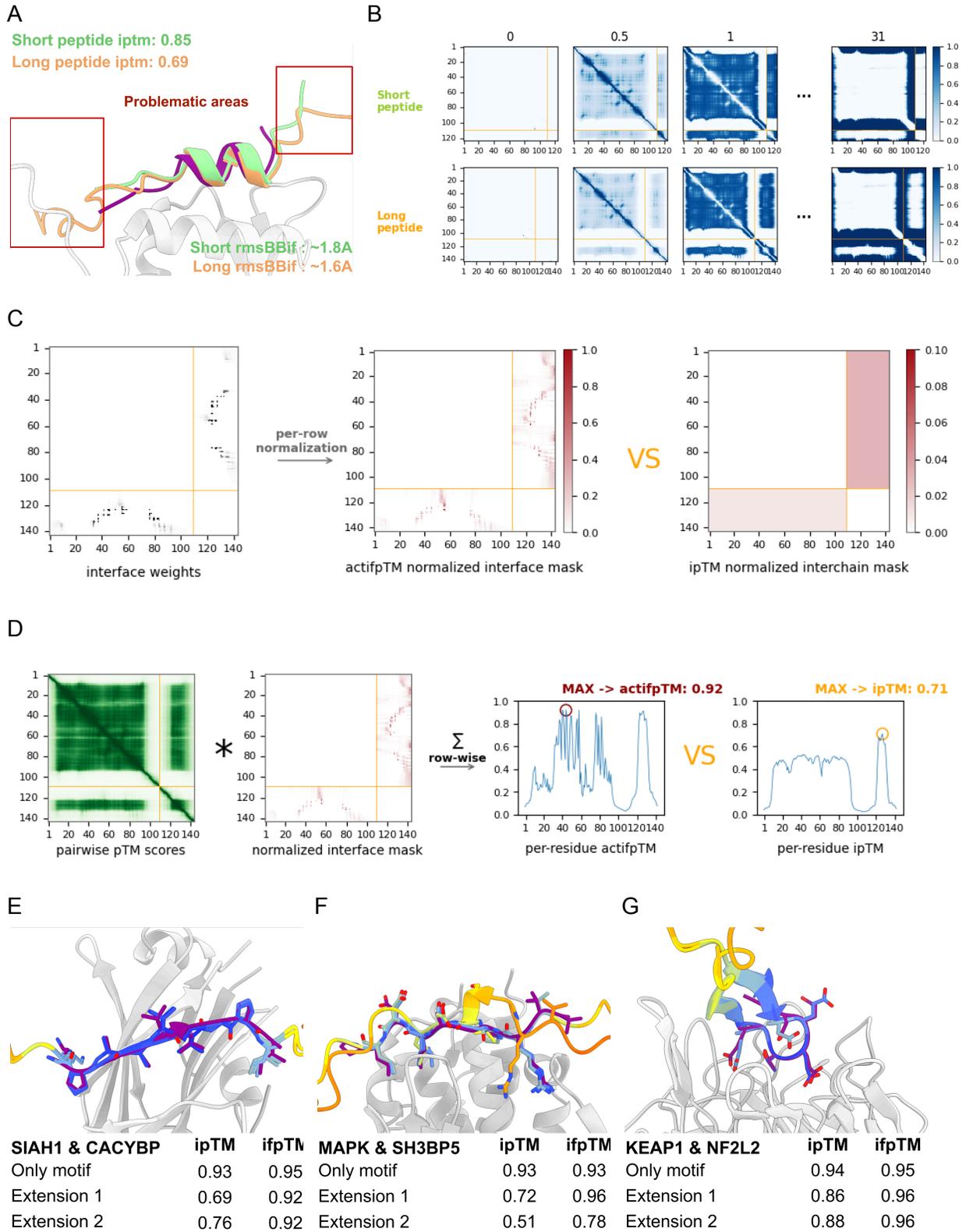

**Figure 1. actifpTM in the AF2 pipeline helps correcting bias from flexible flanking regions. A)** Predictions of MDM2-p53 with p53 peptides of length 14 (15-29) and 34 (5-28) (green and orange, respectively; the native peptide is colored in dark magenta, PDB ID: 1YCR (Kussie *et al.*, 1996)) demonstrate the decrease in ipTM for the longer peptide, even though it is slightly more accurate than the short one (rmsBB_if: RMSD across peptide interface backbone atoms). **B)** The predicted error matrices for each error bin (the first three and the last error bins are shown out of 64) are very similar for the short and long peptide predictions, except for the high error for the flanking regions present only in the longer peptide (white regions). This does not support the large drop in ipTM value upon elongation of the peptide. **C)** In the last steps, the pairwise pTM-score matrix (see **Supplementary Figure 1** for calculation) is multiplied by the actifpTM normalized residue weights, then summed for every residue (row-wise). **D)** As with the original TM-score, the maximum of the values along the sequence is selected as the final actifpTM score. **E-G)** Comparison of ipTM and actifpTM values for example systems. **E)** When the modeled conformation of the defined region of the peptide does not change, actifpTM provides very similar scores, while ipTM demonstrates a considerable decrease (SIAH1 & CACYBP, PDB ID: 2A25 (Santelli *et al.*, 2005)). **F)** actifpTM provides a similar score for similar predictions but drops appropriately upon less confident predictions (MAPK & SH3BP5, PDB ID: 4H3B (Laughlin *et al.*, 2012)). **G)** actifpTM tracks slight changes in conformation more appropriately than ipTM, by not taking into account residues outside the interface. Upon extensions, the peptide prediction became more confident by introducing a beta-hairpin structure with the flanking regions. Unlike actifpTM, ipTM completely misses the overall increase in prediction confidence (KEAP1 & NF2L2, PDB ID: 3ZGC (Hörer *et al.*, 2013)). On all panels, dark magenta denotes experimentally resolved peptides, and modeled peptides are colored according to AF2 pLDDT.

## Results

**Figure 1A** shows a representative result of an AF2 prediction of a peptide-receptor interaction: the model of p53 binding to MDM2. The prediction was performed with two different lengths of the p53 peptide (short: 15-29, long: 5-28). Even though the prediction is of comparable accuracy for the binding motif (both around 1.7 Å RMSD of the peptide interface residue backbone atoms), and pLDDT confidence predictions are similar (see coloring scheme), ipTM values differ dramatically. This phenomenon is a recurring issue when assessing peptide-protein models.

### Implementation of the modified score

This motivated us to implement a score which is not biased by such flexible flanking regions. The main difference between calculating ipTM and actifpTM is the residue weights used in the calculations (**Figure 1C** left *vs.* right). Originally, residue weights are set to 1 for the whole interchain areas, then normalized across rows (**Supplementary Figure 1C**). Instead, we used the contact probability map output by AF2, and calculated the probabilities of each residue-pair (defined here as Cβ-Cβ within 8 Å, Cα is taken for glycines). The resulting pseudo-contact map is in good agreement with the contact map derived from the solved structures (**Supplementary Figure 2**). Next, we follow the usual pTM-score calculation: the residue-weights are row-wise normalized (**Figure 1D** center), and the theoretical maximum TM-scores for each error bin are calculated, similarly to the original TM-score (**Supplementary Figure 1C**). These values are multiplied by their respective error matrices and summed for each row. The maximum of these per-residue actifpTM scores is taken as the final actifpTM score (**Figure 1D**).

We modified the original AF2 code to accommodate these calculations. It was integrated both in the widely-used ColabFold (hence also localcolabfold) pipeline (Mirdita *et al.*, 2022), and can be invoked by checking the *calc_extra_ptm* checkbox in the Jupyter notebook when running on Colab (**Supplementary Figure 3A**), or by using the *--calc_extra_ptm* flag when running localcolabfold. Unlike the one ipTM value calculated across all chains that is provided in the original implementation, we provide in addition actifpTM values for every pair of chains. In the implementation, the pairwise ipTM and per-chain pTM values are also reported and plotted together with the pairwise actifpTM (**Supplementary Figure 3B-D**).

## Case-studies using actifpTM

The implemented code was run on selected domain-motif instances of the dataset from Lee *et al.* (Lee *et al.* 2024), where such problems from the flanking regions were detected earlier, comparing results with the minimal binding motif to those including extensions of two different lengths: 1) with the same length as the motif added on either side, or 2) until the next folded domain or end of the protein. Both ipTM and actifpTM were calculated for all predictions. For visualization purposes, the AF2 models with the highest ipTM were selected for all minimal predictions. In the following we describe a few examples that highlight the advantage of the new actifpTM measure.

(1) Peptide of CACYBP bound to SIAH1 complementing its beta-sheet with a short beta-strand (**Figure 1E**): The predictions for the two different types of extensions are very similar (also in terms of pLDDT of the peptide region resolved in the crystal structure). However the ipTM value for the model using the longer peptide extension is dramatically reduced from 0.93 to 0.69. The actifpTM decreases slightly, from 0.95 to 0.92, but this is negligible, and overall still allows for much more accurate assessment of the model than ipTM.

(2) Binding of SH3BP5 peptide with an extended motif to the binding site of MAPK (**Figure 1F**): there is a considerable change in the binding conformation upon a long extension of the flanking regions (**Figure 1F, extension 2**), although the peptide is still located in the binding site. This change in conformation and also in confidence is reflected in the value of actifpTM, dropping from 0.93 to 0.78. The slight increase between the prediction with only the motif and extension 1 is not surprising since it has already been shown that adding flanking regions to predictions often increases model accuracy (Bret *et al.*, 2024; Lee *et al.*, 2024).

(3) Beta hairpin formation in the interaction between the Kelch domain of KEAP1 and NF2L2 (**Figure 1G**): although the minimal version of the motif is already very confident (ipTM and actifpTM 0.94 and 0.95, respectively), extending allows for forming a beta-hairpin. The actifpTM very slightly increases on this change of conformation, which usually makes binding more tight by decreasing the entropy of the peptide, while ipTM drops, as a result of the flanking regions not confidently modeled.

## Discussion

This study implements actifpTM (<u>act</u>ual <u>i</u>nter<u>f</u>ace pTM), a modified version of AF2 ipTM score, to account for the flexible context of the binding region of peptides. As per its original implementation, ipTM suffers from a bias introduced by these regions, since it takes into account full chains instead of the interface between binding partners. We implemented a simple way to overcome this limitation and score the confidence by weighting interactions between two chains based on their confidence. We demonstrate the merit of actifpTM across several examples, and have incorporated it into widely used versions of AF2.

It is important to note that the current approach requires predictions with approximately equal confidence to deliver equal scores, *i.e.*, it cannot make an overall unconfident prediction confident. However, ipTM has been shown to be good at distinguishing between accurate and inaccurate predictions, sometimes even separating binding from non-binding peptides (Teufel *et al.*, 2023). By correcting for unbound shorter or longer disordered regions around the binding residues, we made this confidence metric more unbiased and comparable across predictions.

Previous studies have also approached the challenge of flexible flanking regions, by e.g. considering only confident positions when calculating iPAE (Kruse *et al.*, 2024; Kim *et al.*, 2024), we note that for the definition of confident interface residues an additional cutoff needed to be defined. In addition, they require post-processing and have not been integrated into commonly run pipelines. Finally, several studies have shown that ipTM was better than iPAE at selecting correct binding conformations (Teufel *et al.*, 2023; Lee *et al.*, 2024; Bret *et al.*, 2024).

We expect that this approach will be particularly useful for assessing predictions of peptide-protein interactions, as the penalty of disordered regions around the binding site is the highest in these cases due to the small number of binding residues.

## Code availability

The method is available as part of the ColabFold (Mirdita *et al.*, 2022) (https://github.com/sokrypton/ColabFold) repository, by checking the *calc_extra_ptm* checkbox in the AlphaFold2.ipynb Jupyter notebook when running on Colab, or by using the *--calc_extra_ptm* flag when running localcolabfold.

A Colab notebook to reproduce figures in this paper is available on GitHub (the notebook was run on a T4 GPU, using Python 3.10.12 and CUDA v12.2):
https://doi.org/10.5281/zenodo.14515382

## Acknowledgements

We thank Tomer Tsaban and Nirit Trabelsi for the critical reading of the manuscript.

## References

Ahdritz,G. *et al.* (2024) OpenFold: retraining AlphaFold2 yields new insights into its learning mechanisms and capacity for generalization. *Nat. Methods*, **21**, 1514–1524.

Baek,M. *et al.* (2023) Efficient and accurate prediction of protein structure using RoseTTAFold2. *bioRxiv*.

Bret,H. *et al.* (2024) From interaction networks to interfaces, scanning intrinsically disordered regions using AlphaFold2. *Nat. Commun.*, **15**, 597.

Evans,R. *et al.* (2021) Protein complex prediction with AlphaFold-Multimer. *BioRxiv*.

Hörer,S. *et al.* (2013) Crystal-contact engineering to obtain a crystal form of the Kelch domain of human Keap1 suitable for ligand-soaking experiments. *Acta Crystallogr. Sect F Struct. Biol. Cryst. Commun.*, **69**, 592–596.

Jumper,J. *et al.* (2021) Highly accurate protein structure prediction with AlphaFold. *Nature*, **596**, 583–589.

Kim,A.-R. *et al.* (2024) Enhanced Protein-Protein Interaction Discovery via AlphaFold-Multimer. *BioRxiv*.

Kruse,T. *et al.* (2024) Substrate recognition principles for the PP2A-B55 protein phosphatase. *BioRxiv*.

Kussie,P.H. *et al.* (1996) Structure of the MDM2 oncoprotein bound to the p53 tumor suppressor transactivation domain. *Science*, **274**, 948–953.

Laughlin,J.D. *et al.* (2012) Structural mechanisms of allostery and autoinhibition in JNK family kinases. *Structure*, **20**, 2174–2184.

Lee,C.Y. *et al.* (2024) Systematic discovery of protein interaction interfaces using AlphaFold and experimental validation. *Mol. Syst. Biol.*, **20**, 75–97.

Mirdita,M. *et al.* (2022) ColabFold: making protein folding accessible to all. *Nat. Methods*, **19**, 679–682.

Santelli,E. *et al.* (2005) Structural analysis of Siah1-Siah-interacting protein interactions and insights into the assembly of an E3 ligase multiprotein complex. *J. Biol. Chem.*, **280**, 34278–34287.

Teufel,F. *et al.* (2023) Deorphanizing peptides using structure prediction. *J. Chem. Inf. Model.*, **63**, 2651–2655.

Zhang,Y. and Skolnick,J. (2004) Scoring function for automated assessment of protein structure template quality. *Proteins*, **57**, 702–710.

# Supplementary Figures for

# actifpTM: a refined confidence metric of AlphaFold2 predictions involving flexible regions

Julia K. Varga[1], Sergey Ovchinnikov[2] and Ora Schueler-Furman[1]*

[1]Department of Microbiology and Molecular Genetics, Institute for Biomedical Research Israel-Canada, Faculty of Medicine, The Hebrew University of Jerusalem, Jerusalem 9112001, Israel
[2]Department of Biology, Massachusetts Institute of Technology, Cambridge, MA 02139
John Harvard Distinguished Science Fellowship, Harvard University, Cambridge, MA 02138

* Correspondence: ora.furman-schueler@mail.huji.ac.il

A

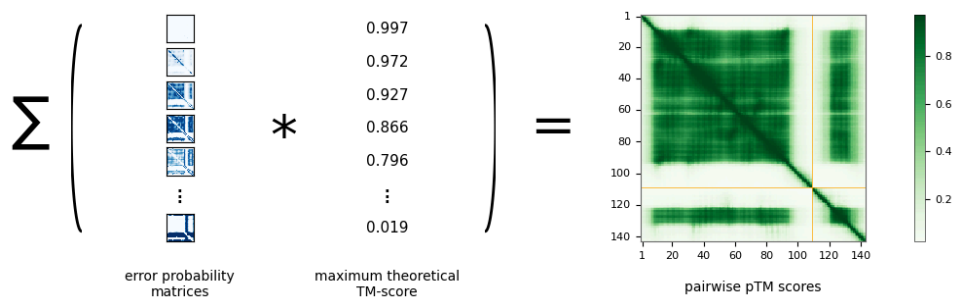

B

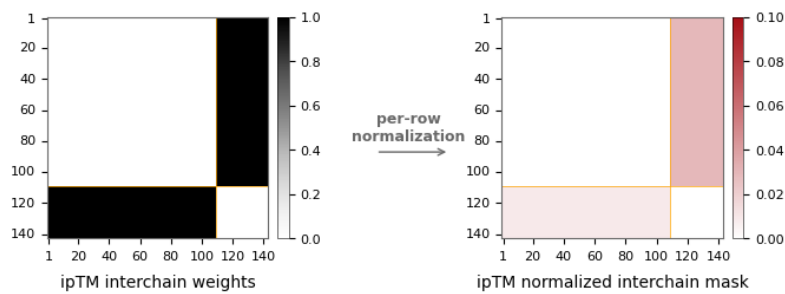

C

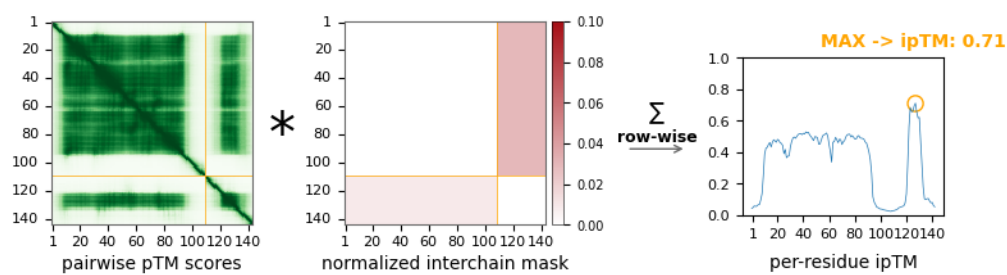

**Supplementary Figure 1. Calculation of ipTM in the AF2 pipeline. A)** A theoretical maximum TM-score is calculated from the values of the error bin centers, then pairwise matrices of each bin are multiplied by the respective theoretical maximum TM-score. Summing them for each pair of residues across bins gives rise to the pairwise TM-score matrix. **B)** The residue weights (equal weights for all residues for pTM and only interchain regions for ipTM calculation) are normalized row-wise. **C)** In the final steps of the algorithm, the pairwise pTM-matrix is multiplied by the normalized pair-residue weights. Then for each aligned residue (Y axis), the values are summarized and the maximum of these values are taken for the pTM or ipTM score.

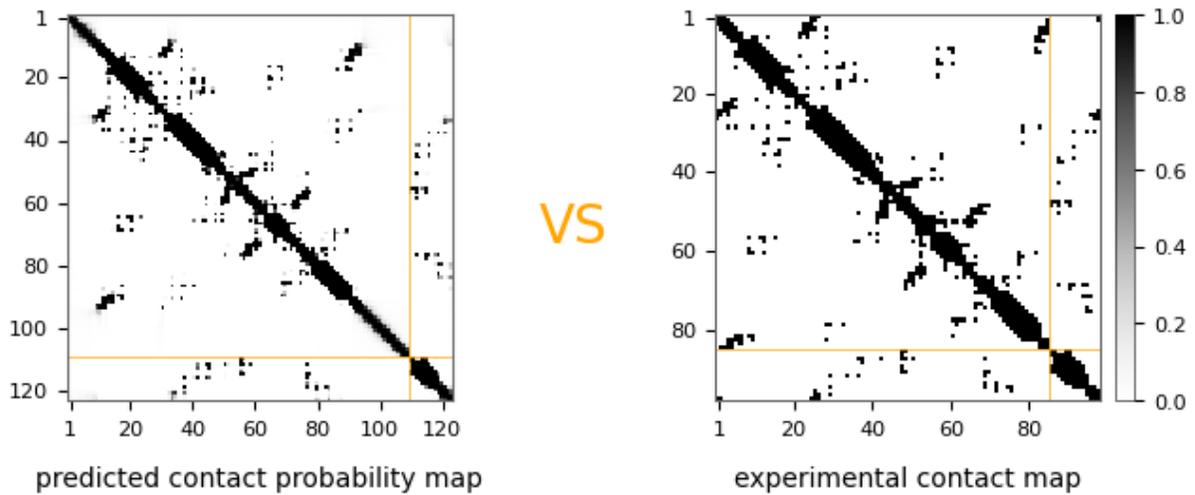

**Supplementary Figure 2. Comparison of contact probabilities and experimental contact map.** Contact probabilities derived from AF2 for a predicted shorter peptide (p53 residues 17-29, left) are in good agreement with the experimental contact map (PDB ID: 1YCR, resolved residues 17-29 in p53, right).

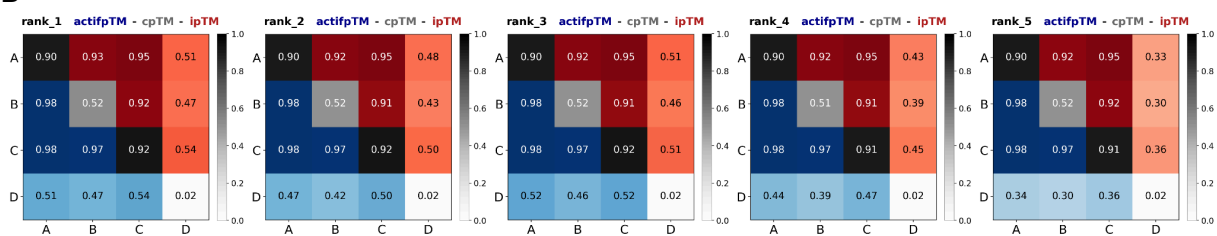

**Supplementary Figure 3. Using ColabFold to calculate actifpTM. A)** Part of the Advanced settings of https://github.com/sokrypton/ColabFold/blob/main/AlphaFold2.ipynb, that now includes the "calc_extra_ptm" option. **B-D)** Example outputs for a run, predicting the complex in PDB ID: 1BHX. **B)** A matrix of pTM values: the lower part of the triangle shows pairwise actifpTM (blue scale), the upper part shows pairwise ipTM (calculated with the standard, interchain method, red scale) and the diagonal values (grey scale) denote chain pTM-s (cpTM). **C)** Pairwise actifpTM, ipTM, chain pTM-s and actifpTM calculated for the full complex are also appended to the output json files of each model **D)** actifpTM of the full complex is also printed in the log at the end of the pipeline.